\documentclass[12pt]{article}
\usepackage[]{psfig}
\usepackage{amsfonts}
\usepackage{epsf}
\newcommand{\be}{\begin{equation}}
\newcommand{\ee}{\end{equation}}
\newcommand{\bea}{\begin{eqnarray}}
\newcommand{\eea}{\end{eqnarray}}
\newcommand{\ba}{\begin{eqnarray}}
\newcommand{\ea}{\end{eqnarray}}
\newcommand{\nn}{\nonumber}

\title{A coordinate-dependent superspace deformation from string theory}
\author{Le\'on G.~Aldrovandi, Fidel A.~Schaposnik\footnote{F.A.S. is associated with CICBA.} and
Guillermo A.~Silva\footnote{G.A.S. is associated with CONICET}\\
\normalsize \it Departamento de F\'\i sica, Facultad de Ciencias
Exactas\\
\normalsize \it
Universidad Nacional de La Plata
}

\begin{document}
%
\date{\today}
\maketitle
\begin{abstract}
Starting from a type II superstring model defined on
$R^{2,2}\times CY_6$ in a linear graviphoton background, we derive
a coordinate
dependent $C$-deformed ${\cal N}=1$, $d=2+2$ superspace. The chiral fermionic coordinates $\theta$
satisfy a Clifford algebra, while the other coordinate algebra
remains unchanged. We find a linear relation between the graviphoton field strength
and the deformation parameter. The null coordinate dependence of the graviphoton background
allows to extend the results to all orders in $\alpha'$.
\end{abstract}

\section{Introduction}
Turning on background fields in  superstring models in the
presence of $D$-branes leads to a deformation of the superspace
geometry. In the case of a constant Neveu-Schwarz B-field
background, the commutator of the space-time coordinates gets
deformed by a central term (see \cite{sw} and  references
therein). If instead one considers a
 Ramond-Ramond background the
odd coordinates of superspace become nonanticommuting
\cite{van}-\cite{seiberg} (for related work see also \cite{t1}-\cite{t3} ).

In the  latter case, if one considers  a type II superstring in
$R^{4}$ with ${\cal N}=2$ supersymmetry in the presence of a
constant self-dual graviphoton field, the effective fermionic
coordinate algebra becomes
\be
\{\theta^\alpha,\theta^\beta\}=C^{\alpha\beta}.
\label{1}
\ee
The connection between the deformation symmetric matrix $C^{\alpha\beta}$
and the constant graviphoton field strength background $F^{\alpha\beta}$
is \cite{van}-\cite{berkovits} and \cite{Billo1}-\cite{Billo2}
\be
C^{\alpha\beta} = {\alpha'}^2 F^{\alpha\beta}.
\label{2}
\ee
This relation is derived using the so-called hybrid formalism
 \cite{berkovits2}-\cite{berko}
which
is particularly adapted to the case of superstrings in Ramond-Ramond
background fields (see also \cite{ccs} for related work).

The graviphoton field strength in (\ref{2}) is selfdual and
constant, therefore not contributing to the energy-momentum tensor
and hence not  backreacting on the flat Minkowski metric.
Moreover, it does not affect the dilaton equations of motion so
that such a background  (flat metric + constant selfdual
graviphoton) is an exact solution of the equations of motion  to
all orders in $\alpha'$.

An interesting possibility is to consider more general graviphoton
backgrounds which could, in principle, lead to a coordinate
dependent deformation $C^{\alpha\beta}(y)$. In particular,
motivated by a suggestion in \cite{shifman}, it was shown in
\cite{ASS} that, within the context of ${\cal N}=1$ $d=4$
euclidean supersymmetry, such a kind of deformation could be
implemented for the Super Yang-Mills model provided the
deformation parameters $C^{\alpha\beta}$ satisfy
\be
 \left.  \bar\sigma^\mu_{\dot\alpha\alpha}\frac{\partial
 C^{\alpha\beta}(y)}{\partial y^\mu}\right\vert_{\theta,\,\bar
 \theta} = 0
   \label{condi}
\ee
with $y^\mu$ the standard hermitian (bosonic) chiral coordinates
defined from the superspace coordinates $(x^\mu,\theta^\alpha,\bar
\theta ^{\dot \alpha})$ as \be
y^\mu=x^\mu+i\theta^\alpha\sigma^\mu_{\alpha\dot\alpha}\bar\theta^{\dot\alpha}
\label{3} \ee As shown in \cite{ASS}, only when condition
(\ref{condi}) is satisfied the (antichiral) superfield strength
$\bar W_{\alpha}$ transforms covariantly under supergauge
transformations and then only in this case a deformed
Super Yang-Mills theory can be consistently defined.

The question that we address in this paper is whether a coordinate
dependent
deformation such that
\be
\{\theta^\alpha,\theta^\beta\}=C^{\alpha\beta}(y)
\label{alg}
\ee
satisfying condition (\ref{condi})
can be obtained in superstring theory
by  considering a D-brane in a non-constant self-dual
background $F^{\mu\nu}(y)$. As we shall see,
by considering a type II  superstring model defined  on
$R^{(2,2)}$  in a family of self-dual graviphoton backgrounds $F_{\mu\nu}(y)$,
we find an affirmative answer to the questions
(the choice of signature will be justified
below).

The paper is organized as follows: in section 2 we present
the worldsheet Lagrangian for a string coupled to a graviphoton
background working in Berkovits hybrid formalism. We write down
the lowest $\alpha'$ order equation
for the graviphoton and, motivated by the pp-wave case, we choose a suitable
class of solutions which allows us
to integrate the equations of motion for the propagators. In section 3 we
obtain the boundary conditions for open strings ending on a D3-brane filling the
$R^{2,2}$ space-time and discuss the supersymmetry preserved.  In section 4
we solve for the propagators and compute the superspace coordinate algebra.
Section 5 contains a discussion of the results.

\section{The worldsheet Lagrangian}

We shall consider a type II  superstring model defined  on
$R^{(2,2)} \times CY_6$ whose ${\cal N}=2$ supersymmetry in
$d=4$  dimensions is deformed by the presence of a self-dual
graviphoton background. Since we  are interested in Ramond-Ramond
backgrounds, it will be convenient to work within Berkovits'  hybrid
formalism \cite{berkovits2}-\cite{berko}. We start from the worldsheet Lagrangian
%
%
\be
{\cal L}_0 = \frac{1}{\alpha'} \left(\frac{1}{2} \tilde
\partial x^\mu \partial x_\mu + p_\alpha\tilde \partial
\theta^\alpha + \bar p_{\dot \alpha} \tilde \partial
{\bar\theta}^{\dot \alpha} + \tilde p_\alpha \partial \tilde
\theta^\alpha + \tilde{\bar p}_{\dot\alpha}
\partial \tilde{\bar \theta}^{\dot \alpha}\right)
\label{s1}
\ee
here $\mu =  1,2,3,4$ and  $\alpha,\dot\alpha = 1,2$ (for metric
and spinor conventions see appendix \ref{spinors}). The bar
denotes space-time chirality while the tilde the worldsheet
chirality. We parametrize the Euclidean signature worldsheet
coordinates with  $z$ and $\tilde z$ and write $\partial =
\partial/\partial z$ and $\tilde
\partial = \partial/\partial \tilde z$. The canonical conjugates
to the complex fermion variables $\theta, \tilde \theta, \bar
\theta$ and $\tilde{\bar \theta}$ are denoted as $p, \tilde p, \bar p$ and
$\tilde{\bar p}$. These conjugate momenta $p$'s can be seen as the
worldsheet versions of $-\partial_\theta|_x, -\partial_{\bar\theta}|_x,
-\partial_{\tilde\theta}|_x$ and $-\partial_{\tilde{\bar\theta}}|_x$.
We have not included
in Lagrangian (\ref{s1}) neither the chiral boson  nor
compactification dependent terms since they will not be relevant
for the following discussion.

Since we are interested in working
in terms of chiral variables $y^\mu$ and derivatives with fixed $y$,
we   change  variables to $y^\mu$, $q_\alpha$, $\tilde q_\alpha$,
$\bar d_{\dot\alpha}$ and  $\tilde{\bar d}_{\dot\alpha}$ given by
\ba
 y^\mu&=&x^\mu+i\theta^\alpha\sigma^\mu_{\alpha\dot\alpha}\bar\theta^{\dot\alpha}+
 i\tilde\theta^\alpha\sigma^\mu_{\alpha\dot\alpha}\tilde{\bar\theta}^{\dot\alpha}\nn\\
 \bar d_{\dot\alpha}&=&\bar p_{\dot\alpha}-i\theta^\alpha\sigma^\mu_{\alpha\dot\alpha}
 \partial x_\mu-\theta\theta\,\partial\bar\theta_{\dot\alpha}+\frac12\bar\theta_{\dot\alpha}
 \partial(\theta\theta)\nn\\
 q_{\alpha}&=&- p_{ \alpha}-i\sigma^\mu_{\alpha\dot\alpha}\bar\theta^{\dot\alpha}
 \partial x_\mu+\frac12\bar\theta\bar\theta\,\partial \theta_{ \alpha}-\frac32
 \partial(\theta_{\alpha}\bar\theta\bar\theta)\nn\\
 \tilde{\!\bar d}_{\dot\alpha}&=&\tilde{\bar p}_{\dot\alpha}-i\tilde\theta^\alpha\sigma^\mu_{\alpha\dot\alpha}
 \tilde\partial x_\mu-\tilde\theta\tilde\theta\,\tilde\partial\tilde{\bar\theta}_{\dot\alpha}+
 \frac12\tilde{\bar\theta}_{\dot\alpha}
 \tilde\partial(\tilde\theta\tilde\theta)\nn\\
 \tilde q_{\alpha}&=&- \tilde p_{ \alpha}-i\sigma^\mu_{\alpha\dot\alpha}\tilde{\bar\theta}^{\dot\alpha}
 \tilde\partial x_\mu+\frac12\tilde{\bar\theta}\tilde{\bar\theta}\,\tilde\partial \tilde\theta_{\alpha}
 -\frac32
 \tilde\partial(\tilde\theta_{\alpha}\tilde{\bar\theta}\tilde{\bar\theta})
\ea
so that  Lagrangian (\ref{s1}) takes  the form
\be {\cal L}_0 = \frac{1}{\alpha'} \left(\frac{1}{2} \tilde
\partial y^\mu
\partial y_\mu
- q_\alpha\tilde \partial \theta^\alpha + \bar d_{\dot \alpha}
\tilde \partial {\bar\theta}^{\dot \alpha}- \tilde q_\alpha
\partial \tilde \theta^\alpha + \tilde{\bar d}_{\dot\alpha}
\partial \tilde{\bar \theta}^{\dot \alpha}\right) \label{s2}
\ee
Here the $ \bar d$'\,s and $q$'\,s replace the $p$'\,s and act
as derivatives with respect to the fermionic coordinates but
keeping $y$ fixed (instead of fixed $x$  as in the $p$'\,s case).
They are the worldsheet versions of the covariant derivative $\bar
D$ and the supercharge $Q$ in target space.

The coupling of the string to an arbitrary graviphoton field background is
implemented by adding to the Lagrangian (\ref{s2}) the vertex
operator for $F_{\mu\nu}$ \be V = \int d^2z\, q_\alpha \tilde
q_\beta F^{\alpha\beta}(y) \label{s3} \ee
where
\be
 F^{\alpha\beta}(y) = \frac{1}{2}\varepsilon^{\alpha\gamma} \left(
 \sigma^{\mu\nu}\right)_\gamma^{\,\beta} F_{\mu\nu}(y).
 \label{s4}
\ee
The total  Lagrangian describing the coupling of the string to a graviphoton background is then
\be {\cal L} = \frac{1}{\alpha'}
\left(\frac{1}{2} \tilde \partial y^\mu
\partial y_\mu
- q_\alpha\tilde \partial \theta^\alpha + \bar d_{\dot \alpha}
\tilde \partial {\bar\theta}^{\dot \alpha}- \tilde q_\alpha
\partial \tilde \theta^\alpha + \tilde{\bar d}_{\dot\alpha}
\partial \tilde{\bar \theta}^{\dot \alpha} + \alpha'
q_\alpha \tilde q_\beta F^{\alpha\beta}(y)\right).
\label{s5}\label{lag}
\ee
As explained in the introduction, in order to get
no back reaction
on the flat metric,  we only
consider the coupling to a self-dual graviphoton background, this
being only possible in  $4+0$ and $2+2$ spaces.

Being  Lagrangian (\ref{lag}) quadratic in $q$ and $\tilde q$ one can easily
integrate them out. Indeed, defining
 \be
  Z[\theta,\tilde \theta; F(y)]\! \equiv \!\!
\int \!\!D\tilde q D q \exp \left(-\frac{1}{\alpha'}\!\!\int\!\! d^2z \left(
-q_\alpha\tilde \partial \theta^\alpha -\tilde q_\alpha
\partial \tilde \theta^\alpha + \alpha' q_\alpha \tilde q_\beta
F^{\alpha\beta}(y) \right) \right) \ee one has
\be Z[\theta,\tilde \theta; F(y)] = \exp\left( -\int d^2 z
 \frac{1}{{\alpha'}^2} F^{-1}_{\alpha\beta}(y)
\partial{\tilde \theta}^\alpha \tilde
\partial \theta^\beta \right) \label{s6} \ee
This leads to an effective Lagrangian of the form
\be {\cal L}_{eff} =  \frac{1}{\alpha'} \left(\frac{1}{2} \tilde
\partial y^\mu
\partial y_\mu + \bar d_{\dot \alpha} \tilde \partial {\bar\theta}^{\dot \alpha}
 +
\tilde{\bar d}_{\dot\alpha} \partial \tilde{\bar \theta}^{\dot
\alpha} + \frac{1}{\alpha'} F^{-1}_{\alpha\beta}(y)
\partial{\tilde \theta}^\alpha \tilde \partial \theta^\beta\right)
\label{leff}
\ee

In a purely bosonic background, the  lowest $\alpha'$
order equation of motion for the graviphoton
field is just the sourceless Maxwell equation
\be
 \partial_\mu F^{\mu\nu} = 0
\label{5}
\ee
This equation is trivially satisfied by
any selfdual field strength which, as explained above, are
precisely those which we consider here. The selfduality condition allows  to write
$ F^{\mu\nu}$ in terms of a $2\times\, 2$ symmetric matrix
$F^{\alpha\beta}$ through the relation
\be
   F^{\mu\nu}(y)=\varepsilon_{\beta\gamma}
   (\sigma^{\mu\nu})_\alpha^{\,\,\gamma} F^{\alpha\beta}(y).
\ee
Being $F^{\mu\nu}$ real,   the
components of $F^{\alpha\beta}$ are also real in $2+2$ dimensions.
It will be useful to rewrite  (\ref{5}) as
\be
   \bar\sigma^\mu_{\dot\alpha\alpha}\partial_\mu F^{\alpha\beta}=0.
   \label{eq2}
\ee
Introducing the null coordinates
$$
 U_1=y_1+ y_4,\;V_1=y_1- y_4\;
$$
\be
 U_2=y_2+ y_3,\;V_2=y_2- y_3,
 \label{lc}
\ee
equations (\ref{eq2}) for the graviphoton field  read
\bea
    \partial_{U_1} F^{11} -  \partial_{V_2} F^{21}=0\; \nonumber\\
    \partial_{U_2} F^{11} +  \partial_{V_1} F^{21}=0\; \nonumber\\
    \partial_{V_2} F^{22} -  \partial_{U_1} F^{12}=0\; \nonumber\\
    \partial_{V_1} F^{22} + \partial_{U_2} F^{12}=0.\label{fmunu}
\eea
The  relevant terms  for our calculations,
considering  the Lagrangian (\ref{leff}) in
terms of the light cone variables (\ref{lc}), are
\bea
 {\cal L}_{eff}^{chiral}= -\frac1{4\alpha'} (\partial U_i \tilde\partial V_i+
 \tilde \partial U_i  \partial V_i)+
 \frac{1}{\alpha'^2}F^{-1}_{\alpha\beta}(U,V)\,
 \partial\tilde \theta^\alpha \tilde \partial \theta^\beta.
\eea
The equations of motion  for the bosonic $(U,V)$ coordinates are
\bea
    \partial\tilde\partial U_i &=& -\frac2{\alpha'} \partial_{V_i}
    F^{-1}_{\alpha \beta} \,\partial\tilde
    \theta^\alpha \tilde \partial \theta^\beta\nonumber\\
     \partial\tilde\partial V_i &=& -\frac2{\alpha'} \partial_{U_i}
    F^{-1}_{\alpha \beta} \,\partial\tilde
    \theta^\alpha \tilde \partial \theta^\beta\label{uv}
\eea

We shall now choose a convenient graviphoton background
satisfying the equations of motion (\ref{fmunu}). With the pp-wave
case in mind \cite{pp},   by taking $F^{12}$ to be a
constant one has from (\ref{fmunu}) that  $F^{11}=F^{11}(V_1,V_2)$ and
$F^{22}=F^{22}(U_1,U_2)$. Then,
choosing also $F^{11}$ to be  constant we obtain from (\ref{uv})
\be
 \Box U_i=0.
\ee
Conformal invariance of the Lagrangian
(\ref{leff}) (see the discussion section) allows one to choose the light cone gauge eliminating the oscillators dependence
of only one of the null $U_i$  coordinates,
we choose this to be $U_1$ getting
\be
 U_1=z+\bar z.
 \label{light}
\ee
Taking $F^{22} = F(U_1)$  we have, in the light cone gauge,  a gaussian Lagrangian.
Let us also note that the choice of $F^{22} = F^{22}(U_1,U_2)$ and constant
$F^{11}$  makes only sense  in $2+2$ signature since in the $4+0$
case $F^{11}$ and $F^{22}$ are related by complex conjugation. Finally, let us
point that the
choice (\ref{light}) is consistent with the boundary conditions (\ref{bbc})
(see below).

Summarizing, our graviphoton
background takes the form
\be
F^{\alpha\beta} = F_0^{\alpha\beta} + \delta^\alpha_2\delta^\beta_2 F(U_1)
\label{background}
\ee
with $F_0^{\alpha\beta}$ a constant symmetric matrix and $F(U_1)$ an arbitrary function.
The choice (\ref{background}) plus conformal invariance results in a gaussian Lagrangian
for the superspace coordinates
in the light cone gauge.

\section{D-brane boundary conditions}

We shall take the worldsheet  ending on a ``D3-brane"
whose worldvolume fills the $2+2$ space. The appropriate boundary conditions for
the open string will be obtained by imposing the cancelation of the boundary terms
in  the equation of motion. When one maps the
disc  to the upper half plane the boundary conditions are
imposed at $z=\tilde z$.
From the Lagrangian (\ref{s5}) one  gets the conditions
\bea
   &&\int_{z=\tilde z} (d\tilde z\, \delta U_i\,\tilde\partial V_i - dz \,\delta
   U_i\,\partial V_i) =0\nn\\
   &&\int_{z=\tilde z} (d\tilde z\, \delta V_i\,\tilde\partial U_i - dz \,\delta
   V_i\,\partial U_i) =0\nn\\
    &&\int_{z=\tilde z} (dz\, q_\alpha\delta\theta^\alpha-
   d\tilde z\,\tilde q_\alpha\delta\tilde\theta^\alpha)=0.
   \label{ferbc}
\eea
We impose Neumann boundary conditions~for the bosonic coordinates of the ``D3-brane"
\be
\left.(\partial-\tilde\partial)U_i\right|_{z=\tilde z}=0,~~~~~~
\left.(\partial-\tilde\partial)V_i\right|_{z=\tilde z}=0.
\label{bbc}
\ee
Concerning the fermionic sector, the cancelation of (\ref{ferbc}) can be achieved by
demanding
\ba
 \left.(q_\alpha-\tilde q_\alpha)\right|_{z=\tilde z}&=&0 \label{tt}\\
 \left.(\theta^\alpha-\tilde \theta^\alpha)\right|_{z=\tilde z}&=&0.
 \label{bc0}
\ea
The equations of motion resulting from Lagrangian (\ref{lag})  take
the form
\ba
 &&\tilde\partial\theta^\alpha-\alpha'F^{\alpha\beta}\tilde q_\beta=0 \nonumber\\
    &&\partial\tilde \theta^\alpha+\alpha'F^{\alpha\beta}q_\beta=0.
\ea
Consistency with (\ref{tt}) demands that
\be
    \left.(\tilde\partial  \theta^\alpha+\partial
    \tilde \theta^\alpha)\right|_{z=\bar z}=0.
    \label{bc}
\ee
The boundary conditions (\ref{tt})-(\ref{bc0}) preserve half
of the the supersymmetries. Indeed, in the absence of D-branes, Lagrangian (\ref{lag})
 is invariant under the transformations
\be
 \delta\theta^\alpha=\epsilon^\alpha,~~~~~~\delta\tilde\theta^\alpha=\tilde\epsilon^\alpha
\ee
where $\epsilon^\alpha$ and $\tilde\epsilon^\alpha$ are constants. These
transformations give rise to  conserved independent
supercharges
\be
 Q_\alpha=\oint  dz\, q_\alpha,~~~~ \tilde Q_\alpha=\oint d\bar z\, \tilde q_\alpha.
\ee
When a D-brane  is present  the expression for the charges is
\be
 Q_\alpha=\int_C  dz\, q_\alpha,~~~~ \tilde Q_\alpha=\int_C d\bar z\, \tilde q_\alpha.
\ee
where $C$ is a semi-circle centered on the origin. Conditions (\ref{tt}) imply
that $Q$ and $\tilde Q$ are equal and no longer independent.

\section{Propagators and (anti)commutation relations}

To obtain the anticommutation relations for the fermionic  coordinates we need
to compute the following propagators,
\bea
   G_1^{\alpha\beta}(z,\tilde z | w,\tilde w ) &\equiv&
   \langle\theta^\alpha(z,\tilde z)\theta^\beta(w,\tilde w)\rangle
   \nonumber\\
   G_2^{\alpha\beta}(z,\tilde z | w,\tilde w)  &\equiv&
   \langle\theta^\alpha(z,\tilde z)\tilde\theta^\beta(w,\tilde w)\rangle \nonumber\\
   G_3^{\alpha\beta}(z,\tilde z | w,\tilde w) &\equiv&
   \langle\tilde\theta^\alpha(z,\tilde z)\tilde\theta^\beta(w,\tilde w)\rangle
\eea
which, according to Lagrangian (\ref{leff}),
should obey the following differential equations
\be
    \partial_z\left( F^{-1}_{\alpha\beta}(z,\tilde z)
    \tilde \partial_{\tilde z} G_i^{\beta\gamma}(z,\tilde z | w,\tilde w)\right)=
    -\alpha'^2\delta_\alpha^\gamma\,\delta^2(z-w)\,\delta_i^2
    \label{eq3}
\ee
Conditions (\ref{bc0}) and (\ref{bc}) at $w=\tilde w$ imply the following relations between the
propagators $G_i$
\bea
    G_1^{\alpha\beta}(z,\tilde z | w,\tilde w)&=&
    G_2^{\alpha\beta}(z,\tilde z | w,\tilde w)\label{uno}\\
    G_2^{\alpha\beta}(w,\tilde w | z,\tilde z)&=&
    G_3^{\alpha\beta}(w,\tilde w |z,\tilde z )\\
    \tilde\partial_{\tilde w} G_1^{\alpha\beta}(z,\tilde z | w,\tilde w)&=&
    -\partial_w G_2^{\alpha\beta}(z,\tilde z | w,\tilde w)\label{dos}\\
    \tilde \partial_{\tilde w} G_2^{\alpha\beta}(w,\tilde w |z,\tilde z )&=&
    -\partial_w G_3^{\alpha\beta}(w,\tilde w | z,\tilde z)
\eea
The Grassmann character of $\theta$ and the conjugation
relations among them (see appendix) impose  the constraints
\bea
    G_1^{\alpha\beta}(z,\tilde z | w,\tilde w)&=&
    -G_1^{\beta\alpha}(w,\tilde w | z,\tilde z)\label{tres}\\
    G_3^{\alpha\beta}(z,\tilde z | w,\tilde w)&=&
    -G_3^{\beta\alpha}(w,\tilde w | z,\tilde z)\\
    (G_1^{\alpha\beta}(z,\tilde z | w,\tilde w))^*&=&
    G_3^{\beta\alpha}(w,\tilde w |z,\tilde z )\label{5'}\\
    (G_2^{\alpha\beta}(z,\tilde z | w,\tilde w))^*&=&
    G_2^{\beta\alpha}(w,\tilde w |z,\tilde z)\label{cuatro}
\eea

In order to go further, we shall consider the following
  graviphoton background
\be
 F^{\alpha\beta}(U,V)=F^{\alpha\beta}_0 + \lambda\,
U_1\, \delta^\alpha_2 \delta^\beta_2.
\label{linbackground}
\ee
Using the conformal invariance
of the action one can fix  $U_1=z+\tilde z$ (more on
the conformal invariance of the action in the discussion section). For this choice
only the $(22)$ component of the propagators computed in
\cite{van}-\cite{berkovits} for a constant background is modified.
Moreover, since boundary conditions (\ref{bc0}) and (\ref{bc})
are preserved under complex conjugation  one needs
only to solve for $G_1$  and  $G_2$, this is because due to the
complex conjugation properties in $2+2$   $G_3$ can be obtained as the
complex conjugate of $G_1$ (see eqn.(\ref{5'})). The propagators satisfying the
boundary conditions (\ref{bc0}) and (\ref{bc}) are
\bea
    G_1^{\alpha\beta}(z,\tilde z | w,\tilde w)&=&-\frac{\alpha'^2}{2\pi}
    \left(F^{\alpha\beta}(z,w)\ln\frac{\tilde z-w}{z-\tilde
    w}+\lambda(\tilde z -z + w -\tilde w)\delta^\alpha_2 \delta^\beta_2\right)\nonumber\\
    G_2^{\alpha\beta}(z,\tilde z | w,\tilde w)&=&-\frac{\alpha'^2}{2\pi}
    \left(F^{\alpha\beta}(z,\tilde w)\ln\frac{|z-w|^2}{(z-\tilde
    w)^2}+\lambda(\tilde z -z + w -\tilde w)\delta^\alpha_2 \delta^\beta_2\right)\nonumber\\
    G_3^{\alpha\beta}(z,\tilde z | w,\tilde w)&=&-\frac{\alpha'^2}{2\pi}
    \left(F^{\alpha\beta}(\tilde z,\tilde w)\ln\frac{\tilde z-w}{z-\tilde
    w}+\lambda(\tilde z -z + w -\tilde w)\delta^\alpha_2 \delta^\beta_2\right)
\nonumber\\
\eea
where
\be
F^{\alpha\beta}(z,w)=F^{\alpha\beta}_0 + \lambda (z+w)
\delta^\alpha_2 \delta^\beta_2.
\ee
Taking $z$ and $w$ to be on the boundary $z=\tilde z=\tau,\;w=\tilde
w=\tau' $ one gets
\be
    \langle\theta^\alpha(\tau)\theta^\beta(\tau')\rangle=\frac {i}{2}\alpha'^2
    F^{\alpha\beta}(\tau,\tau')\,{\rm sgn}(\tau-\tau')
\ee
from which one can get that the $\theta$  anticommutator as
\ba
   \{\theta^\alpha,\theta^\beta\}&=&\lim_{\epsilon\to 0} \left(\theta^\alpha(\tau+\epsilon) \theta^\beta(\tau)+
   \theta^\beta(\tau ) \theta^\alpha(\tau-\epsilon)\right)\\
   &=&i\alpha'^2(F^{\alpha\beta}_0 + \lambda U_1
   \delta^\alpha_2 \delta^\beta_2)
\ea
or
\be
\{\theta^\alpha,\theta^\beta\}=i\alpha'^2F^{\alpha\beta}(y).
\ee
Hence, as in the case of a constant background, the deformation
of the $\theta$ anticommutator   is proportional to the
graviphoton field strength. From  (\ref{5}) one immediately
verifies that the deformation parameter satisfies the condition
(\ref{condi}) found in \cite{ASS} by demanding consistency of the
gauge theory.
Although we have obtained this result
for  a
field strength of the form  (\ref{linbackground}),
we expect that an analogous result should hold  for more general cases.
Also, since the $\bar\theta$ coordinates were not affected
by the background coupling, the commutation relations
of $\bar \theta$ with  $y^\mu$ and $\theta$  are not modified.
The same happens with the commutator between $\theta$ and $y^\mu$

Concerning the commutation relations for  bosonic space-time
coordinates among themselves, the relevant propagators to consider are
\bea
    K_1^{ij}(z,\tilde z | w,\tilde w) &=&
    \langle U_i(z,\tilde z)U_j(w,\tilde w)\rangle\nonumber\\
    K_2^{ij}(z,\tilde z | w,\tilde w) &=&
    \langle U_i(z,\tilde z)V_j(w,\tilde w)\rangle\nonumber\\
    K_3^{ij}(z,\tilde z | w,\tilde w) &=&
    \langle V_i(z,\tilde z)V_j(w,\tilde w)\rangle
\eea
which, according to Lagrangian (\ref{leff}) should obey the following equations
\bea
    \Box_z K_1^{ij}(z,\tilde z |
    w,\tilde w) &=& 0\nonumber\\
    \Box_z  K_2^{ij}(z,\tilde z | w,\tilde w)
    &=& 2\alpha'\delta^{ij}\delta^2(z-w) \nonumber\\
    \Box_w K_2^{ij}(z,\tilde z | w,\tilde w)
    &=& 2\alpha'\delta^{ij}\delta^2(z-w)-\nonumber\\
    &&
    \frac{2}{\alpha'}\langle U_i(z,\tilde z)\,\partial_{U_j} F^{-1}_{\alpha\beta}(w,\tilde w)\,
    \partial\tilde\theta^\alpha(w,\tilde w)\tilde\partial\theta^\beta(w,\tilde w) \rangle\nonumber\\
    \Box_z K_3^{ij}(z,\tilde z | w,\tilde w)
 &=& \frac{2}{\alpha'}\langle\partial_{U_i} F^{-1}_{\alpha\beta}(z,\tilde z)
    \tilde\partial\theta^\alpha(z,\tilde z)\partial\tilde\theta^\beta(z,\tilde z)\,
    V_j(w,\tilde w)\rangle
    \label{bp}
\eea
here $\Box_z=\partial_z \tilde\partial_{\tilde z}$.
Divergent $\delta(0)$ terms arise in
(\ref{bp}) from
tadpole graph when contracting
derivatives of   $\theta$ fields in the last two lines. They are put to zero by an appropriate
regularization. Once this is done, equations  (\ref{bp})
are replaced by
\bea
    \partial_z \tilde\partial_{\tilde z}K_1^{ij}(z,\tilde z |
    w,\tilde w) &=& 0\nonumber\\
    \partial_z \tilde\partial_{\tilde z}K_2^{ij}(z,\tilde z | w,\tilde
    w) &=& 2\alpha'\delta^{ij}\delta^2(z-w) \nonumber\\
    \partial_z \tilde\partial_{\tilde z}K_2^{ij}(w,\tilde
    w | z,\tilde z) &=&\ 2\alpha'\delta^{ij}\delta^2(z-w)\nonumber\\
    \partial_z \tilde\partial_{\tilde z}K_3^{ij}(z,\tilde z | w,\tilde
    w)
 &=& 0
\eea
which are just the equations one obtains in the constant $F_{\mu\nu}$ case
leading to trivial commutation relations for the chiral coordinates $y^\mu$
 \cite{seiberg},
 so that also in our coordinate dependent background one has
\be
[y^\mu,y^\nu] = 0
\ee

\section{Discussion}

Starting from a type II superstring model defined on
$R^{2,2}\times CY_6$ we have constructed
a deformed ${\cal N}=1$, $d=2+2$ superspace with a coordinate
dependent deformation.
Indeed, by turning on a self dual linear
graviphoton background and using Berkovits hybrid
formalism we have been able to compute the propagators for the
coordinates of  superspace and from them infer the deformed algebra of the
supercoodinates.

The resulting deformation is of the same type as that
proposed in \cite{ASS}:  the chiral fermionic
coordinates $\theta$ are not Grassman variables but satisfy a
Clifford algebra of the type
$\{\theta^\alpha,\theta^\beta\}=C^{\alpha\beta}(y)$ while the
other coordinates, $y,\; \bar\theta$, remain (anti)commuting.
 We find that a linear relation between the
deformation parameter and the graviphoton field strength
$C^{\alpha\beta}(y) = {\alpha'}^2 F^{\alpha\beta}(y)$ holds, as
in the case of a constant background \cite{van}-\cite{seiberg}.
This relation implies that
the deformation parameter satisfies the condition
$\partial_\mu C^{\mu\nu}=0$, also required in order to define
a consistent Super Yang-Mills theory in such a deformed superspace
 \cite{ASS}.

The results above were obtained to the lowest $\alpha'$ order. In
order to extend its validity to all orders, one should address
the problem of conformal
invariance of the theory defined by  Lagrangian (\ref{lag}). Now,
one can  easily see that conformal invariance
is guaranteed
to all orders. Indeed,  our choice of background is
independent of the null coordinates $V_1,\,V_2$
(or, alternatively, $U_1,U_2$). Then, when  one   splits the
$U_i$ fields in the form $U_i = U_i^b + U_i^q$,
with $U_i^b$ a classical background  and $U_i^q$ the quantum
fluctuations,
this last has nothing to contract with and hence
it does not contribute to the effective action. Then, after setting
$U_i= U_i^b$, one ends with a   Lagrangian (\ref{lag})  quadratic in the
quantum fields. It is then easy to check that this quadratic Lagrangian has
no conformal anomaly\footnote{We thank N.~Berkovits for explaining us this
point.}.

Concerning the deformed superalgebra we obtained, the
following  comments are worth mentioning. First, note this algebra for the
supercoordinates implies that the ordinary bosonic coordinates
$x=y-i\theta\sigma\bar\theta$ are noncommutatives and its
deformation parameter is also coordinate dependent. Also, it is
interesting to note that the resulting deformation can be
easily generalized. Indeed, as shown in \cite{van}, a nonvanishing
commutator between the bosonic chiral coordinates $y$ can be obtained by
turning on a NS-NS two form $B$ while a similar non trivial result
is gotten for the $y-\theta$ commutator through the gravitini
$\Psi$. Finally, note that the linear dependence of the $C$
parameter with the space-time coordinate is very similar to an old
proposal of Schwarz and van Nieuwenhuizen \cite{van nieu} where
they analyze a possible substructure of the space-time through the
relation
$\{\theta^\alpha,\theta^\beta\}=\gamma_\mu^{\alpha\beta}x^\mu$.

The study of a coordinate dependent $C$-deformation, started in \cite{ASS}
and continued in the present work was prompted by the observation in \cite{shifman}
on the connection between this deformation and the spectral degeneracy in
SUSY gluodynamics. Now that we have obtained such deformation starting from a superstring
theory, this connection should be more thoroughly studied.
We hope to report on this issue in a
forthcoming work.

\section*{Appendix: Spinors in 2+2}
\label{spinors}

In $d=2+2$ a Majorana-Weyl (real and chiral matrix being diagonal)
representation for the gammas is \be
 \Gamma^\mu=\left( \begin{array}{cc}
   0 & \sigma^\mu \\
   \bar \sigma^\mu & 0 \\
 \end{array}\right),~~~
 \sigma^\mu=\left(1,i\sigma^2,\sigma^1,\sigma^3\right),~~
 \bar \sigma^\mu=\left(-1,i\sigma^2,\sigma^1,\sigma^3\right)
 \label{rep}
\ee where $\mu,\nu= 1,2,3,4$ and the metric is
$\eta=(--++)$\footnote{Note that in $2+2$ space all $\sigma^\mu$ are real, so
$(\sigma^\mu)^*=\sigma^\mu$ (cf. with the Lorentzian case where
the $\sigma^\mu$ are Hermitic, this is
$(\sigma^\mu)^*=(\sigma^\mu)^T$). No simple relation exist in
Euclidean space.}.

It then follows that \be
 C_{ab}=\left( \begin{array}{cc}
   -\epsilon^{\alpha\beta} & 0 \\
   0 & -\epsilon_{\dot\alpha\dot\beta} \\
 \end{array}\right),~~~\epsilon^{\alpha\beta}=\epsilon_{\dot\alpha\dot\beta}=
 \left( \begin{array}{cc}
   0 & 1 \\
   -1 & 0 \\
 \end{array}\right)=i\sigma^2
\ee The inverse spinor metrics are defined as \ba
 \epsilon_{\alpha\beta}\epsilon^{\beta\gamma}&=&\delta_\alpha^\gamma
 \Longrightarrow\epsilon_{\alpha\beta}=-\epsilon^{\alpha\beta}\\
 \epsilon_{\dot\alpha\dot\beta}\epsilon^{\dot\beta\dot\gamma}&=&\delta_{\dot\alpha}^{\dot\gamma}
 \Longrightarrow\epsilon_{\dot\alpha\dot\beta}=-\epsilon^{\dot\alpha\dot\beta}
\ea which we use to define \be
 C^{ab}=\left( \begin{array}{cc}
   \epsilon_{\alpha\beta} & 0 \\
   0 & \epsilon^{\dot\alpha\dot\beta} \\
 \end{array}\right),~~~\epsilon_{\alpha\beta}=\epsilon^{\dot\alpha\dot\beta}=
 \left( \begin{array}{cc}
   0 & -1 \\
   1 & 0 \\
 \end{array}\right)=-i\sigma^2
\ee Raising and lowering of indices are defined as \be
 \psi^\alpha=\epsilon^{\alpha\beta}\psi_\beta,~~\psi_\alpha=\epsilon_{\alpha\beta}\psi^\beta
\ee The index structure for  $\sigma^\mu,\,\bar \sigma^\mu$ is \be
 \sigma^\mu_{~\alpha\dot\alpha},~~~~\bar \sigma^{\mu\,\dot\alpha\alpha}
\ee and moreover they are related by rising/lowering indices as
\be
 \sigma^{\mu~\alpha\dot\alpha}=\epsilon^{\alpha\beta}\epsilon^{\dot\alpha\dot\beta}
 \sigma^\mu_{~\beta\dot\beta}=\bar \sigma^{\mu\,\dot\alpha\alpha}\,.
\ee Since in $2+2$ the matrices
$\sigma^{\mu\nu}=\frac14(\sigma^\mu\bar\sigma^\nu-\sigma^\nu\bar\sigma^\mu)$
are real,   we conclude that $\chi_\alpha$ and $(\chi_\alpha)^*$
transform in the same way under $Spin(2,2)$. This can be
seen as the reason for the existence of Majorana-Weyl spinors in
2+2. It is then consistent to impose the condition
$(\chi_\alpha)^*=\chi_\alpha$. Complex conjugation in $2+ 2$  does
not change chirality (as in $3+1$) neither it changes the index
position (as in $4+0$). When acting on a product of spinors
we define it to invert  their order
\be
 (\theta\chi)^*=(\theta^\alpha\chi_\alpha)^*=(\chi_\alpha)^*(\theta^\alpha)^*
 =\chi_\alpha\theta^\alpha=-\theta^\alpha\chi_\alpha=-\theta\chi
\ee \be
 (\theta\sigma^\mu\bar\chi)^*=(\theta^\alpha\sigma^\mu_{~\alpha\dot\alpha}\bar\chi^{\dot\alpha})^*
 =(\bar\chi^{\dot\alpha})^*(\theta^\alpha)^*\sigma^\mu_{~\alpha\dot\alpha}
 =-\theta\sigma^\mu\bar\chi\label{current}
\ee in these relations we have taken the spinors to be MW  and
used that the $\sigma^\mu$ matrices are real. Although working
with real spinors,  it is $i\theta\chi$ and
$i\theta\sigma^\mu\bar\chi$ that are real.

The conjugation properties of the variables used  in the text, which correspond to a pair
of MW spinors or just simply one complex Weyl spinor, are
\be
 (\theta_\alpha)^*=\tilde \theta_\alpha,~~~~(p_\alpha)^*=-\tilde p_\alpha
\ee Similar relations hold for dotted indices.

\vspace{1.5 cm}

\noindent\underline{Acknowledgements}: We wish to thank N.~Berkovits
for his continuous help and suggestions. We also acknowledge
O.~Chand\'\i a and C.~N\'u\~nez for helpful comments. L.A. is supported
by CONICET. This work is partially supported by UNLP, CICBA, and
CONICET.

\end{document}